\documentclass[aps,pra,
epsfigure,twocolumn,longbibliography,superscriptaddress]{revtex4-1}

\usepackage{amsfonts}
\usepackage{amssymb}
\usepackage{amsmath}
\usepackage{calc}
\usepackage{subfigure}
\usepackage{graphicx}
\usepackage{epstopdf}
\usepackage{dcolumn}
\usepackage{bm}
\usepackage{color} 
\usepackage{txfonts}
\usepackage[dvipsnames]{xcolor}
\usepackage{appendix}
\usepackage[colorlinks, citecolor=blue]{hyperref} 

\newcommand{\beq}{\begin{equation}}
	\newcommand{\eeq}{\end{equation}}
\newcommand{\beqa}{\begin{eqnarray}}
	\newcommand{\eeqa}{\end{eqnarray}}

\newcommand{\new}[1]{\textcolor{magenta}{#1}}

\begin{document}
	
	\title{Effective Scaling Approach to Frictionless Quantum Quenches in Trapped Bose Gases}
	\date{\today}
	
	\author{Tang-You Huang}
	\email{huangtangyou@shu.edu.cn}
   \affiliation{International Center of Quantum Artificial Intelligence for Science and Technology (QuArtist) \\ and Department of Physics, Shanghai University, 200444 Shanghai, China}
   
   \affiliation{Department of Physical Chemistry, University of the Basque Country UPV/EHU, Apartado 644, 48080 Bilbao, Spain}
	
	\author{Michele Modugno}
	\email{michele.modugno@ehu.eus}
   
    \affiliation{Department of Physics,
	University of the Basque Country UPV/EHU, 48080 Bilbao, Spain}
   \affiliation{IKERBASQUE, Basque Foundation for Science, 48013 Bilbao, Spain}

	\author{Xi Chen}
	\email{chenxi1979cn@gmail.com} 	
	\affiliation{Department of Physical Chemistry, University of the Basque Country UPV/EHU, Apartado 644, 48080 Bilbao, Spain}
%	\affiliation{International Center of Quantum Artificial Intelligence for Science and Technology (QuArtist) \\ and Department of Physics, Shanghai University, 200444 Shanghai, China}

\begin{abstract}

%In this paper, 
We work out the effective scaling approach to frictionless quantum quenches 
in a one-dimensional Bose gas trapped in a harmonic trap. The effective scaling approach produces an auxiliary equation for the scaling parameter interpolating between {the}
noninteracting and the Thomas-Fermi limits.  This allows us 
to implement {a} frictionless quench 
by engineering inversely the smooth trap frequency, as compared to the two-jump trajectory. Our result is beneficial to design the shortcut-to-adiabaticity expansion of trapped Bose gases for arbitrary values of interaction, and can be directly extended to the three-dimensional case.

\end{abstract}

\maketitle

\section{introduction}

Bose-Einstein condensates (BECs) and their related phenomena - such as collective excitations, collapse, and nolinear
dynamics, to mention a few - have aroused great interest
since their first experimental realization  \cite{nobel2002rmp,stringari1999RMP}.
% From {the} theoretical point of view, weakly interacting BECs obeys the Gross-Pitaevskii (GP) equation
 %in the mean-field approximation 
From the theoretical
point of view, weakly interacting BECs can be accurately
described within the framework of the Gross-Pitaevskii
(GP) theory, which provides a remarkable agreement
% provides an accurate description  of {the} dynamics, which is qualitative agreement 
with experimental observations.
In most cases of classical hydrodynamics \cite{Stringari1996prl} or scaling transformations \cite{Castin1996prl,KaganPRA96}, % are focused on to solve analytically the
exact analytical solution can be found for the collective dynamics and free expansion of BECs in time-dependent harmonic traps, both in the noninteracting limit and the Thomas-Fermi (TF) regime \cite{stringari1999RMP}.  
In this vein,  symmetries give birth to an intriguing property of self-similarity, which allows to utilize  the scaling approach for describing the dynamics of ultracold atomic systems, for instance, the atomic gases in the non-interacting and the hydrodynamic regimes \cite{GritsevNPJ2010,ModugnoPRA2011}, Tonks-Girardeau (TG) gas of impenetrable bosons \cite{Stringari2003praTG,TGPRL05},  superfluid Fermi  gas \cite{ModugnoPRA2011,FermiPRA2010},
and thermal cloud \cite{BrunnPRA2000} in different geometries. 

Besides, an effective scaling approach has further been proposed as an approximate solution for the evolution of both bosonic and fermionic density distributions, for describing the collective dynamics of a trapped Bose gas \cite{odelin2002pra}, the expansion of Fermi gas  \cite{Stringari2002prl,FermiBSC-BECPRA06}  and of quantum degenerate Bose-Fermi mixtures \cite{MichelePRA2003}.
%Since there doesn't exist exact self-similarity in the crossover between the mean-field TF regime and the TG regime, a strongly interaction one-dimensional Bose gas does not follow the scaling solution  \cite{Stringari2003praTG,OhbergPRL02}. In order to complement it,  this effective scaling, consisting in a self-similar evolution in the hydrodynamic regime to be satisfied on average by integrating over the spatial coordinates, reduce the complexity of numerical treatment. Recently, the accuracy of such effective approach has been proved in reproducing the exact solution of one-dimensional (1D) and three-dimensional (3D) GP equations for arbitrary values of the interactions \cite{MichelePRA2018,Michele2020PRR}.   
It consists in a self-similar evolution
in the hydrodynamic regime to be satisfied on average by integrating
over the spatial coordinates, reduce the complexity
of numerical treatment. Recently, the accuracy of such effective
approach in reproducing the exact solution of quasi
one-dimensional (1D) and three-dimensional (3D) GP equations
for arbitrary values of the interactions has been discussed
in Refs. \cite{OhbergPRL02,MichelePRA2018}. Remarkably, it turns out that the space-averaged
self-similarity can provide an accurate description
in several situations \cite{Michele2020PRR}.

%Lieb-Liniger states \cite{Sutherland2002prl,Lieb-Liniger2008prl}.
%Remarkably, rely on the property of self-similarity, the quench-induced breathing mode \cite{breath-model1,quench-breath-model} and the quench-based coherent manipulation\cite{adolf2011pra} of ultra-cold atom have been formulated.
%It's paved an alternative strategy for learning and controlling the dynamic of interacting quantum gas in terms of self-similarity.

In a slightly different but relevant topic,  the concept of shortcuts to adiabaticity (STA), originally proposed for fast schemes reproducing or approaching slow adiabatic process \cite{TORRONTEGUI2013117,staRev}, have extended further the control paradigms for frictionless atomic cooling in a expanding harmonic trap \cite{Muga2009frictionless,ChenPRL104,SchaffPRA2010fast,SchaffEPL2011,del2011fast,rohringer2015non}. In the context of inverse engineering,  the scaling approach \cite{Muga2009frictionless} and Lewis-Riesenfeld dynamical invariant \cite{ChenPRL104} bring out the various forms of the Ermakov equation for the scaling parameter, capturing the character of the self-similar evolution. Along with it, the harmonic trap frequency is thus inversely engineered for the propose, by choosing  an interpolation function of the scaling parameter with the appropriate boundary conditions. This strategy can be applicable to other ultracold
atomic systems as a TG gas \cite{adolfoTG}, an anisotropic gas containing quantum defects \cite{stringari} and  a Fermi gas \cite{haibinpra}. However, tracking back to a BEC described by the GP equation in the mean-field approximation,  one can realize that the original Ermakov equation obtained in non-interacting case needs to be modified in the TF limit or in the case of a time-dependent interaction \cite{Muga2009frictionless,rohringer2015non} . To remedy it, the variational approximation \cite{zoller1996prl} (which is equivalent to moment method \cite{juan1999prl}),  can be complemented  by the concept of STA, for studying the dynamics of BECs \cite{jingSciRep,tangyou2020chaos,tangyou2020pra}, valid for the range from zero to small atomic interaction, with the implication on the quantum speed limits and quantum thermodynamics \cite{jing2018njp,tianniu2020prr}. As a matter of fact, the accuracy of the variational approximation depends on the presumed ansatz in terms of nonlinearity \cite{tangyou2020chaos,tangyou2020pra}.  Thus, the motivation of this work is to fill the gap in more general theory on STA design for 1D Bose gas with arbitrary interactions.

In this paper, we integrate the effective scaling approach into inverse engineering for frictionless quantum quenches in trapped Bose gas, for arbitrary values of the atomic interaction strength. 
%with the atomic interaction varying in the whole range of the values between zero and infinity. 
Here we focus on the 1D GP equation, but the result can be extended to 3D case. By assuming the scaling solution in the hydrodynamic limit, we derive the Ermakov-like equation for the scaling parameter interpolating between non-interacting and the TF limit. With this, the frictionless quench is designed, and also
compared to the free expansion and two-jump trajectory of STA. Finally, the numerical simulation is performed to check the stability of our method, 
and the energetic cost of STA is discussed as well.

%In this paper, the self-similar dynamic of atomic interacting Bose gas subject to the GP equation, has been investigated by an effective scaling approach (ESA) \cite{michele2018pra,Michele2020prr} based on hydrodynamic theory. 
%A second-order differential Newton-liked equation has been produced for the effective scaling factor which characterizes the width of the wave-packet, and the validity of such equation verified by a quenching process.  
%And the phenomenological analytical solution of quench process has been formulated with promising accuracy.
%Moreover, the free expansion and breathing oscillation of the wave-packet has been predicted, and numerically proved in section B.
%In section C, we have combined the STA for frictionless engineering the system by reversely manipulating the trap frequency. We found the STA engineering of interacting BECs maintains high fidelity even for the arbitrary interacting regime. 

\section{Effective scaling approach}

\label{secII}

We start by considering {a} {quasi-}1D BEC confined in cigar-shaped trap, {characterized by a longitudinal frequency $\omega_{0}$ and a tight transverse frequency $\omega_{\perp}\gg\omega_{0}$. Therefore, the system can be effectively described by a wave function $\psi(x,t)$,}
whose dynamics is governed by the following GP equation, 
\beq\label{gp}
i \frac{\partial \psi }{\partial t}=-\frac{1}{2}\frac{%
	\partial ^{2} \psi }{\partial x^{2}}+\frac{1}{2} \omega
^{2}(t)x^2 \psi+g |\psi |^{2}\psi,
\eeq
{that is written here in dimensionless form, for convenience. To this end, we have used $l_0= \sqrt{\hbar/m \omega_0}$ as unit length ($m$ being the particle mass), $\hbar\omega_{0}$ as unit energy, and $\omega_{0}^{-1}$ as unit time. The interaction strength can be written in terms of the scattering length $a_{s}$ as $g=2Na_{s}/\ell_{0}$, with $N$ being the number of atoms and with the total density being normalized to one.}
%Here all the quantities are dimensionless, for 
%convenience, by introducing $t = \omega_0 \tilde{t}$, $\omega(t)= \tilde{\omega}(t)/\omega_{0}$, $x= \tilde{x}/l_0$, $g= \tilde{g}/ \hbar \omega_{0} l_0$,
%where $m$ is atomic mass, $\omega_{0} \equiv \tilde{\omega}(0)$ is the initial longitudinal trapping frequency, $l_0= \sqrt{\hbar/m \omega_0}$ is the corresponding cloud size, $\tilde{g}=2N \hbar a_{s} \omega_{\perp} $ with the scattering length $a_s$, atom number $N$ and transversal trapping frequency $\omega_{\perp}$.  
%\TY{The quantities with tilde are the original correspondences with the physical units in the GP equation, which is omitted here for simplicity. }

In order to elaborate the effective scaling approach, we apply the Madelung transformation $\psi=\sqrt{n(x,t)}e^{i\phi(x,t)}$, such that the Lagrangian {of the system can be written as}
%is presented as 
\beq
\label{Lagrangian_density}
\mathcal{L}=- \left[\partial_t\phi+\frac{1}{2}(\bigtriangledown\phi)^2+ \frac{1}{8}\left(\frac{\bigtriangledown n}{n}\right)^2
+ V(x,t) +gn\right]n,
\eeq
where $V(x,t)= \omega
^{2}(t)x^2/2$. 
The essence of {the} variational Lagrangian formalism is to minimize {the action} $S= {\iint} \mathcal{L} dxdt$ with respect to the parameters $q_i =\{n,\phi\}$, that is,  $\delta S/\delta q_i = 0$. 
%The condition for such minimization gives 
{The latter corresponds to} the  Eular-Lagrangian equations
%, 
$\partial \mathcal{L}/ \partial q_i- d(\partial \mathcal{L}/\partial \dot{q}_i )/dx=0$, from which the hydrodynamic equations are obtained as \cite{stringari1999RMP}
%,
\beqa
\label{continuity}
\frac{\partial n}{\partial t} + \frac{\partial\left(n\bigtriangledown\phi\right)}{\partial x} = 0, 
\\
\frac{\partial (\bigtriangledown\phi)}{\partial t}+\partial_x\left( P(x,t)+ \frac{1}{2}v^2+V(x,t)+gn\right)=0,
\label{eular}
\eeqa
where  
%$P \left(x,t\right) = -\frac{1}{2\sqrt{n}}\partial^2_x\sqrt{n}$ 
{$P \left(x,t\right) = -(\partial^2_x\sqrt{n})/(2\sqrt{n})$}
is the 
{so-called} 
quantum pressure.
By inserting the scaling solution 
%\cite{stringari1999RMP},
$n = n_0(x/a)/a$ and $v = \dot{a}x/a$ \cite{stringari1999RMP} into Eq. (\ref{eular}), one can obtain 
\beq
-\frac{1}{2}\frac{\ddot{a}}{a}x^2 = P(x,t) +V(x,t)+ gn -P(0,t)-gn(0,0).
\eeq
Multiplying the resulting equation by $n_0(\xi)$ after rescaling the spatial as $\xi = x/a$, and integrating over 
%$d\xi$
{the coordinates}, we get the following {\textit{effective}} Ermakov-like equation \cite{MichelePRA2018}
\beq
\label{GEE}
\ddot{a} + \omega^2(t)a = \frac{A}{a^3}+\frac{B}{a^2},
\eeq
where $A$ and $B$ read 
\beqa
\label{AandB}
A = \frac{P(0,0)-E_k^0}{E_v^0}, ~~B = \frac{gn(0,0)-E_{int}^0}{E_v^0},
\eeqa
{with $E_{k}^{0}\equiv(1/2)\int[\partial_{x}\sqrt{n_{0}(x)}]^{2}dx$ being the kinetic energy, 
$E_{v}^{0} \equiv \int V(x,0)n_{0}(x)dx$ the potential energy, and $E_{int}^{0}\equiv(g/2)\int n_{0}^{2}(x)dx$ the interaction energy, all evaluated at the initial time  $t=0$. Here $n_{0}(x)\equiv n(x,0)$ represents the ground state of the system in the harmonic trap. As shown in Ref. \cite{MichelePRA2018}, the parameters $A$ and $B$ satisfy the relation  $A+B=1$ for arbitrary (positive) interactions.}

%with $E_k^0=$, $E_{int}^0$, and  $E_v^0$  being  the kinetic,
%interaction, and potential energies at $t=0$, \new{and} $E^0_{tot} = E_{k}^0 +E_{int}^0 +E_{v}^0$ being the total energy.  To obtain the coefficients $A$ and $B$, we need calculate the kinetic, interaction and potential energy, $E_{k}(t) = \int[\partial_{x}\sqrt{n(x,t)}]^2/2dx$, 
%the interaction energy is $E_{int} (t)= g\int n^2(x,t)dx$,
%and the potential energy is $
%E_{v}(t) = \int V(x,t)n(x,t)dx$ at $t=0$.

%\begin{figure}[t]
%	\includegraphics[width=\columnwidth]{fig1.eps}
%	\caption{ The parameter $A$, $B$ and relation $A+B$, as functions of arbitrary repulsive  interaction $g$ calculated by $(\ref{AandB})$ represented by red dashed, blue solid, and black dot-dashed lines, respectively. Here we employ the time-imaginary method to obtain $n(x,0)$, the initial eigenstate of trapped Bose gas with the interaction covering  the whole range from weak to ultra-strong interacting regime.}
%	\label{fig1}
%\end{figure}

{Remarkably, the Ermakov-like Eq. (\ref{GEE}) permits to describe the dynamics of the system in terms of an effective self-similar evolution, for arbitrary interactions. Note that this equation is more accurate than the one obtained by applying a Gaussian ansatz for arbitrary interactions, $\ddot{a} + \omega^2(t)a = 1/a^3+ g/(\sqrt{2\pi} a^2)$ \cite{zoller1996prl,tangyou2020chaos}.
Obviously, Eq. (\ref{GEE}) reproduces the exact scaling in the non-interacting ($g=0$) and Thomas-Fermi limits ($g \gg 1$). Namely, for $g=0$ } 
%
%In general, the Ermakov-like equation (\ref{GEE}) captures the dynamics arbitrary interactions.  
%
%%Firstly, it is easy to conclude that 
%%the parameters $A$ and $B$ satisfy the relation  $A+B = 1$ 
%%with arbitrary (positive) interaction \cite{MichelePRA2018}, 
%%by taking into account   { $E_{tot}^0 = \mu(0)$ and $\mu(0)= P(0,0)+gn_0(0,0)$}, 
%%Fig. \ref{fig1} illustrates the values of $A$ and $B$ as the function of atomic interaction $g$ subject to repulsive interaction, 
%%where $n(x,0)=|\psi(x,0)|^2$ is calculated from the initial eigenstate by the imaginary-time method.  
%
%Secondly,  it is evident that the Ermakov-like equation (\ref{GEE}) covers two aforementioned cases for non-interacting  atoms and BEC in the TF limit, where negligible ($g \simeq 0$) or very strong ($g \to +\infty$) interactions are assumed. 
%Obviously, in the non-interacting case $g=0$, 
the above Eq. (\ref{GEE}) 
%reduces 
{corresponds} 
to the original Ermakov equation,
\beq
\label{Ermakov}
\ddot{a} + \omega^2(t)a = 
%\frac{1}{a^3}
{1/a^{3}},
\eeq 
%when the Gaussian ansatz $n_0= (1/\pi a^2)^{1/2} e^{-x^2/a^2}$ is chosen, yielding $A=1$ and $B=0$. 
{with $n_0(x)= e^{-x^2/a^2}/\sqrt{\pi a^2}$ and $A=1$, $B=0$.}
This is consistent with the results derived from variational control \cite{zoller1996prl,tangyou2020chaos} and also from {the} scaling approach and Lewis-Riesenfeld dynamical invariant \cite{ChenPRL104,Muga2009frictionless}. 
{In the opposite TF regime, the ground state density is $n_{0} =[\mu-\omega(0)x^2/2]/g$, with $\mu=E_{int}^0  +E_{v}^0$ being the chemical potential (the kinetic energy can be safely neglected in this limit \cite{stringari1999RMP}. Then, inserting $n_{0}$ } 
%On the contrary, in the TF approximation, the initial density is chosen as $n_{0} =[\mu-\omega(0)x^2/2]/g$, with chemical potential $\mu=E_{int}^0  +E_{v}^0$, by neglecting the kinetic energy. Inserting this 
into Eq. (\ref{AandB}) 
%produces
{yields ($A=0$, $B=1$)} 
\beq
\label{ErmakovTF}
\ddot{a} + \omega^2(t)a = 
%\frac{1}{a^2}
{1/a^2},
\eeq 
%with $A=0$ and $B=1$, from which we cover the result in TF limit by the scaling approach 
{which corresponds to the exact TF result} 
\cite{Castin1996prl,KaganPRA96,tangyou2020chaos}. 

%In addition, we should emphasize that , Eq. (\ref{GEE}) has more accuracy than the one, 
%\beq
%\ddot{a} + \omega^2(t)a = \frac{1}{a^3}+ \frac{g}{\sqrt{2\pi} a^2}.
%\eeq
%derived from variational approximation  for weakly interacting atoms with Gaussian ansatz \cite{zoller1996prl,tangyou2020chaos},  though the interaction $g$ is invoked. 
%This implies $A=1$ and $B=g/\sqrt{2 \pi}$, which does not fulfil $A+B=1$ for non-negligible interaction $g$, and are not equal to $A=1$
%and $B= 4(\sqrt{2}-1) g/\sqrt{2 \pi}$ when the Gaussian asantz is used directly in Eq. (\ref{AandB}). 

\section{Shortcuts to Adiabaticity}

In this section, we use {the} Ermakov-like equation (\ref{GEE}) to construct {a} STA protocol for $\omega(t)$, for achieving {a} frictionless quench from 
%an 
{the} 
initial trap frequency 
%$\omega(0)=\omega_i$ 
{$\omega_i\equiv\omega(0)$ (with $\omega(0)=1$ fixed by our notation choice)} 
to a final 
%one $\omega(t_f)=\omega_f$ 
{value $\omega_f\equiv\omega(t_f)$} 
within a short time $t_f$, 
%when  $\omega_f<\omega_0$
{with $\gamma\equiv\sqrt{\omega_i/\omega_f}>1$. { That is,
in a finite-time non-adiabatic expansion,	the trap frequency is changed to some lower final value, while keeping the populations of the initial and final levels invariant, thus without generating friction and heating.}
% there is no final excition heat here frictionless means... [IT COULD BE HELPFUL TO EXPLAIN IT]}. 
The  frictionless cooling of ultracold atoms trapped in time-dependent harmonic traps
%has been firstly 
{was originally} 
investigated in two noninteracting and TF limits \cite{Muga2009frictionless,ChenPRL104}. Later, the variational approximation 
%is utilized 
{was used} 
to design 
%such 
{the same} 
process in 
{the} 
weakly interacting 
%case 
{regime} 
\cite{tangyou2020chaos}.   Here, we propose
%concentrate on the general method for the whole regime of atom interaction ranging from zero to infinity values.
{a general approach based on the formulation discussed in previous section, for arbitrary values of the interactions,
ranging from the non-interacting to the TF regime.}

\begin{figure}[tbp]
 \includegraphics[width=\columnwidth]{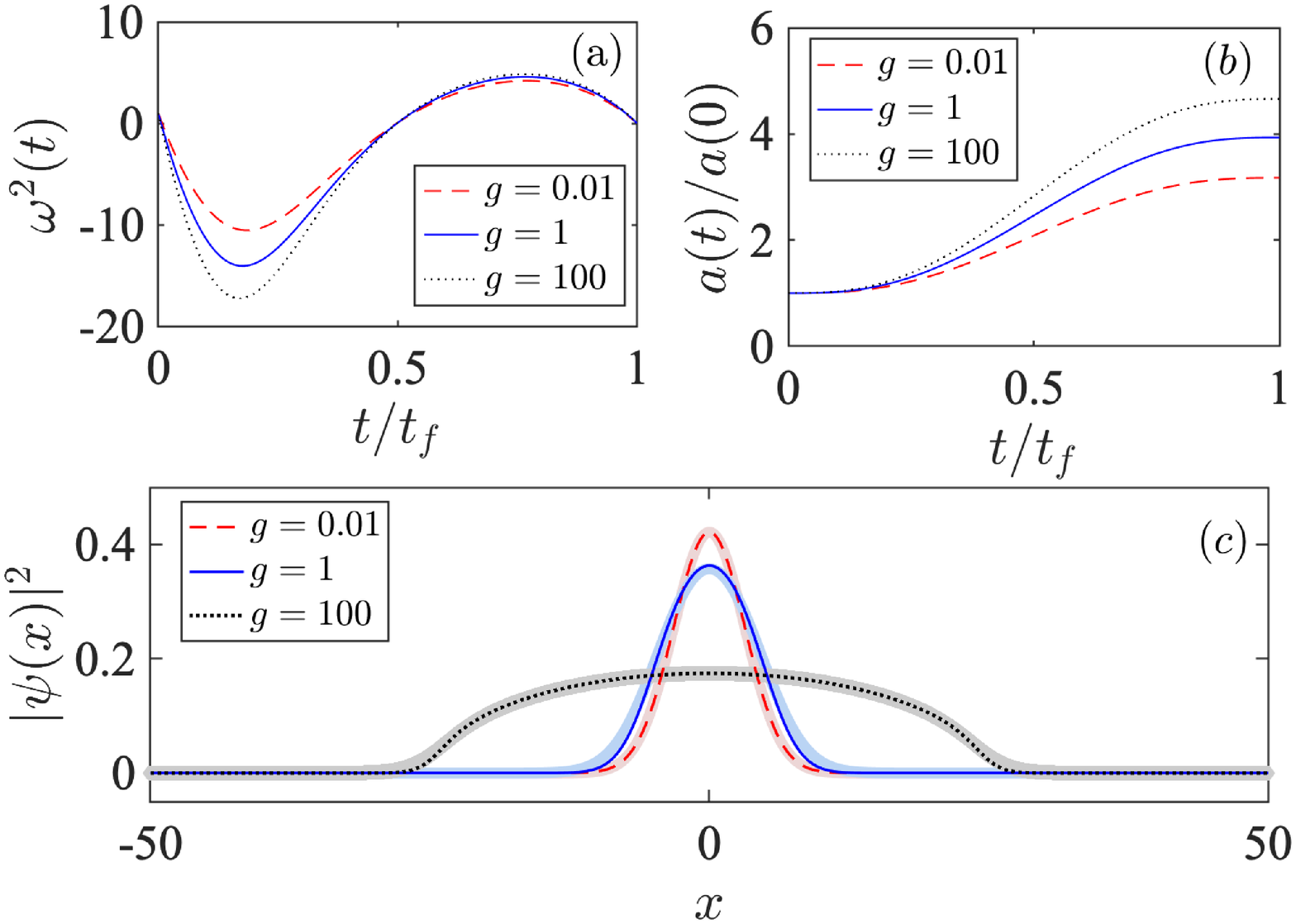}
 \caption{The designed trap frequency $\omega^2(t)$ (a) and {the} corresponding width $a(t)$ (b) as a function of time, for 
 % illustrated in (a)  with 
 $g = 0.01$ (red-dashed {line}), $g = 1$ (blue solid {line}) and $g = 100$ (black dashed {line}).  (c) The final states (thick lines)  are compared to the corresponding stationary states 
 %(wide curves)
 {(thin lines)}.  Parameters: 
 %$\omega_i=1$, 
 $\omega_{f}=0.1$ and  $t_f = 1$.}
	\label{sta}
\end{figure}

To this end, we recast Eq. (\ref{GEE}) 
%to 
{in the form of} 
the perturbative Kepler problem in classical mechanics, 
%and obtain 
{in the presence of} 
the effective potential 
\beq
\label{potential}
\mathcal{U}(a) = \frac{A}{2a^2} +\frac{B}{a}  + \frac{1}{2}\omega^2(t)a^2,
\eeq
for {a} fictitious  particle with unit mass, satisfying the Newton equation $\ddot{a} = - \partial \mathcal{U}(a)/\partial a$, derived from Eq. (\ref{GEE}). 
%As a consequence, the 
{The} total energy of {the} particle reads 
\beq
\label{totalE}
E(a) = \frac{\dot{a}\new{^{2}} }{2}+ \frac{A}{2a^2} +\frac{B}{a}  + \frac{1}{2}\omega^2(t)a^2.
\eeq 
%from which the conditions $\dot{a}=0$ and $\partial \mathcal{U}(a)/\partial a=0$ for minimizing the energy are suggested to achieve the  adiabatic evolution \cite{jingSciRep}, obeying
{The conditions for an adiabatic evolution are $\dot{a}=0$ and $\partial \mathcal{U}(a)/\partial a=0$ \cite{jingSciRep}, yielding} 
\beq
\label{Reference}
a^4 \omega^2 (t)- a B = A.
\eeq
%\TY{Defining} 
{Then, we define} the time-averaged energy %as follows 
\beq
\label{energycost}
\mathcal{E} = \frac{1}{t_f} \int_{0}^{t_f} E(a) dt, 
\eeq
{that will be used} for quantifying the energetic cost of STA, in 
%for 
the discussion below. 
%By assuming that the initial trap frequency $\omega_i = 1$, and final one $\omega_f = 1/\gamma^2$, i.e. $\gamma=\sqrt{\omega_i/\omega_f}$,  one has to impose the following boundary conditions:
{The initial boundary conditions read \cite{jingSciRep}:} 
\beqa
\label{Bounday condition-1}
a(0) = a_{\mathrm{i}},  ~ \dot{a}(0) = \ddot{a}(0)=0,
\\
\label{Bounday condition-2}
a(t_f) = a_{\mathrm{f}}, ~ \dot{a}(t_f) = \ddot{a}(t_f)=0,
\eeqa
where $a_{\mathrm{i}}$ and $a_{\mathrm{f}}$ are the unique positive real solutions of Eq. (\ref{Reference}), at $t=0$ and $t=t_f$.
%\beqa
%a_{\mathrm{i}}^4-Ba_{\mathrm{i}} = A, \label{ai}
%\\
%\frac{1}{\gamma^4}a_{\mathrm{f}}^4-Ba_{\mathrm{f}} = A\label{af}.
%\eeqa
These boundary conditions (\ref{Bounday condition-1}) and (\ref{Bounday condition-2}) guarantee that the initial and final states are adiabatic correspondences for designing STA protocols. 
%With the fixed
{Having fixed the} boundary conditions, the trajectory of $a(t)$
can be interpolated, by choosing 
%simplest 
{a simple} 
polynomial ansatz 
%as
{of the form}  
\beq
\label{polyansatz}
a(t)=a_{\mathrm{i}}-6(a_{\mathrm{i}}-a_{\mathrm{f}})s^{5}+15(a_{\mathrm{i}}-a_{\mathrm{f}})s^{4}-10(a_{\mathrm{i}}-a_{\mathrm{f}})s^3,  
\eeq%
with $s=t/t_f$. Consequently, the trap frequency $\omega(t)$ is determined by Eq. $(\ref{GEE})$. If an imaginary trap frequency is allowed, the harmonic trap inverts to a parabolic repeller, instead of 
%the 
{a} trap, such that $t_f$ may be formally made arbitrarily short. However, 
%the physical constraints, i.e. anharmonicity and laser intensity,  always exist in practice, and $|\omega^2(t)|\leq \delta$, sets the minimal time, 
{since in experimental implementations there are always imperfections and limitations related, e.g., to the trap anharmonicity and to the laser power, this poses a constraint of the amplitude of the frequency that can be physically achieved, namely $|\omega^2(t)|\leq \delta$}, with  $\delta$ being a real number \cite{tangyou2020chaos,Stefanatos2010pra,Stefanatos2012pra}.

%Figure \ref{sta}
{In Fig. \ref{sta} we show the results for a STA protocol of a frictionless quench from $\omega_{i}=1$ to $\omega_{f}=0.1$, within a time $t_f = 1$. Panels} (a) and (b) illustrate the 
%STA protocols of 
{the evolution of the} 
trap frequency $\omega(t)$ and 
%the corresponding time evolutions 
of {the} width $a(t)$ 
%with 
{for} 
different {values of the} interactions, $g = 0.01, 1, 100$.
%, where the parameters $\omega_i=1$, $\omega_{f}=0.1$, and $t_f = 1$ are used. 
With the designed protocol, 
we use the split-operator method to 
%obtain 
{propagate numerically the wave function to} 
the final state  $|\psi_f\rangle $, that is then compared to the ground state of the final trap $|\tilde{\psi}_f\rangle$, see Fig. \ref{sta}(c) (the latter is computed by means of a standard by imaginary-time evolution).} 
% numerically, and further compare the stationary one  $|\tilde{\psi}_f\rangle $, calculated by imaginary-time method in Fig. \ref{sta} (c).  Moreover, the dependence of fidelity $F= |\langle \tilde{\psi}_f (x) |\psi(x,t_f)\rangle|^2$ on the atomic interaction at differential final trap frequencies $\omega_{f}= 0.2, 0.1, 0.05$  are depicted in Fig. \ref{fidelity}. 
%Clearly, the influence of the value of interaction on the STA trajectory is pronounced, but the fidelities of STA  are still reasonable in the whole range of the values between zero and infinity. However,  the fidelity is slightly lower 
%
{The corresponding fidelity, $F\equiv|\langle \tilde{\psi}_f |\psi_{f}\rangle|^2$, is shown in Fig. \ref{fidelity} 
as a function of the interaction strength $g$, for different values of the trap frequency.  Notice that $F\equiv1$ in both the non-interacting ($g\ll1$) and TF limits ($g\gg1$), where the scaling ansatz is an exact solution.
Remarkably, the fidelity always stays very close to one even in the intermediate regime, thanks to the accuracy of the effective scaling approach \cite{MichelePRA2018}. There the deviations from $F=1$ are less than $1\%$, and show a weak dependence on the trap frequency, i.e., the fidelity decreases $\omega_{f}$ is decreased.} 
 %Clearly, the influence of the value of interaction on the STA trajectory is pronounced, but the fidelities of STA  are still reasonable in the whole range of the values between zero and infinity. However,  the fidelity is slightly lower 
 %\new{It slightly decreases} in the intermediate regime, since there exist exact scaling solution in the non-interacting and TF limit, and  the scaling solution in the hydrodynamical regime is applied in the effective scaling approach. 
 %Also the fidelity is decreased as well when the trap frequency is smaller.  
 This is due to the fact that the interaction becomes dominant when the trap frequency is negligible.
 %, in which case that the sech-type soliton has more accuracy as an appropriate ansatz \cite{tangyou2020pra} \new{[BUT THIS WAS FOR A SOLITON; HERE WE HAVE NO SOLITONS...]} . 
 %The property of
 {Notice also that the behaviour of the}  fidelity presented here can be also intuitively understood 
 %by 
 {in terms of} the stability of {a} particle %staying at 
 {in the presence of} the effective potential, see Eq. (\ref{potential}). 

 \begin{figure}[tbp]
	\includegraphics[width=\columnwidth]{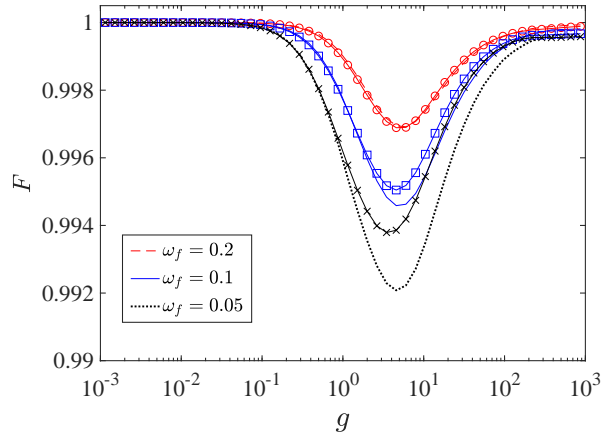}
	\caption{ 
	The fidelity {$F$ (see text)} versus the %atomic 
	interaction {strength} $g$, for a frictionless quench (symbols) and a sudden quench (lines). %[PLEASE ADD ALL SYMBOLS IN THE LEGEND]
	Three final frequencies are considered: $\omega_f = 0.05,0.1,0.2$. Notice that, whereas in the first case the final time is fixed to $t_{f}=1$, in case of sudden quench it depends on the (final) frequency, see Eq. (\ref{bangt}).
	%, where two  STA protocols, designed from smooth polynomial ansatz and two-jump trajectory, are compared. Parameters: 
	%$\omega_i=1$, 
%	for polynomial ansatz we choose $t_f=1$ for different frequencies $\omega_f = 0.2$ (dashed red), $0.1$ (solid blue) and $0.05$ (dotted line), and for two-jump trajectory
%	$\omega_f = 0.2$ (dashed red with ``$\circ$"), $0.1$ (solid blue with ``$\square$") and $0.05$ (dotted line with ``$\times$") corresponding to different $t_f$, calculated from Eq. (\ref{bangt}). 
}	 
	\label{fidelity}
\end{figure}

\section{Sudden Quench}

For completeness, 
%we shall study the quench-induced free expansion and breathing modes based on Eq. $(\ref{GEE})$. To compare the STA protocol, we describe the sudden quench by the following piecewise function:
{here we also consider the case of a sudden quench of the trap frequency:}
\beq
\label{quench_func}
\omega(t)= 
\begin{cases}
~	\omega_i, \qquad  t = 0\\
~	\omega_f,  \qquad  0<t < t_f
\end{cases}.
\eeq
In the non-interacting case, the conventional Ermakov equation (\ref{Ermakov}) gives the analytical solution 
\beq 
\label{VA_solution}
a(t) = \sqrt{1+(\omega_i^2-\omega_f^2)\sin^2(\omega_f t)/\omega_f^2},
\eeq
with initial boundary conditions $a(0) =1$, and $\dot{a}(0)=0$,
which 
%describes the oscillation of collective dynamics with the 
{corresponds to a collective oscillation with} 
period  $\tau = \pi/\omega_f$. {Therefore, in general the boundary conditions (\ref{Bounday condition-1}) and (\ref{Bounday condition-2}) cannot be not attained at $t=t_{f}$, and this implies a heating/excitation of the system.}

\begin{figure}[tbp]
	\includegraphics[width=\columnwidth]{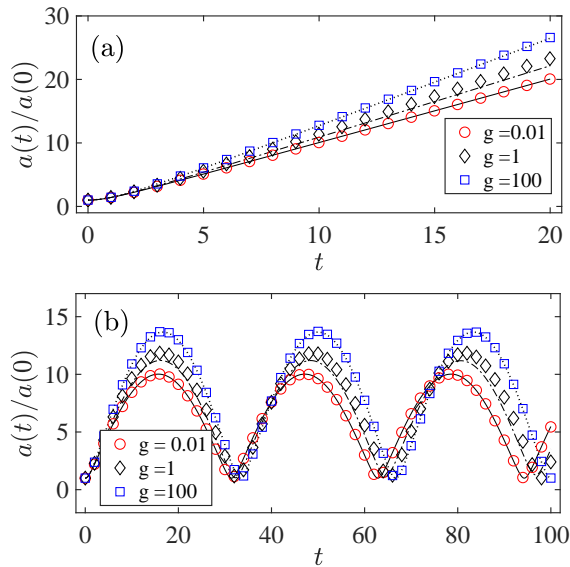}
	\caption{ (a) Dynamics of free expansion when the trap is suddenly switched off, i.e. $\omega_f = 0$.  (b) Collective oscillation modes after quenching trap frequency to $\omega_f = 0. 1$.  For comparison,  the center of the mass of the wave packet, $a(t)= \sqrt{\langle x^2 \rangle- \langle x \rangle^2}$ , calculated from numerical simulation, is denoted by symbol. Parameters: $g = 0.01$ (sold line; with red circle), $g = 1$ (dashed line, with black diamond), and $g = 100$ (dotted line, with red square),	and the initial trap frequency is $\omega_0 = 1$. }
	\label{aaa}
\end{figure}

In the limit $\omega_f \to 0$, {the case of a free expansion,} Eq. (\ref{VA_solution}) 
%is reduced 
{reduces} 
to $a(t) = (1+\omega_i^2 t^2)^{1/2}$.
In more general cases, 
%the 
analytical solutions 
%of 
{for a } sudden quenches are not available, but one can 
%solve it numerically by using Eq. (\ref{GEE}).
{still solve Eq. (\ref{GEE}) numerically.} 
Fig. \ref{aaa}, 
{we demonstrate that  Eq. (\ref{GEE}) is accurate enough to describe the dynamics in sudden quenches, when the values of nonlinear interaction is changed from zero to infinity.  
} 
%{DO WE SHOW ANYTHING ABOUT THIS POINT?} 
%From Eq. (\ref{VA_solution}), we see that the breathing modes are induced by sudden quench from $\omega_i$ to $\omega_{f} \neq 1$ at $t=0$. 
%This implies the heating or excitation since the boundary conditions (\ref{Bounday condition-1}) and (\ref{Bounday condition-2}) are not attained.  

%However, the reminiscence of sudden quench allows us to reproduce the STA protocol with simple two jumps, as proposed firstly for the compression of solitons in nonlinear fibers \cite{Anderson94}. 
{Based on the above considerations, we can build a STA protocol with just two quenches, as proposed for the compression of solitons in nonlinear fibers \cite{Anderson94}: an initial quench from $\omega_{i}$ to $\omega_{c}$ at $t=0$, and a second one from $\omega_{c}$ to $\omega_{f}$, at the final time $t_{f}$ such that} 
\begin{equation}
	t_{f}=\pi /\left( 2\omega _{c}\right),  
	\label{T}
\end{equation}
{with $\omega _{c}=\sqrt{\omega_i\omega_f}$. Then, in the non-interacting case we have [see again Eqs. (\ref{quench_func}) and (\ref{VA_solution})]} 
%Following \new{Refs.} \cite{chen2010transient,tangyou2020chaos},  we begin with the non-interacting limit $g \simeq 0$ to obtain the analytical solution as a heuristic case study. In this scenario, the trap frequency begins with $\omega_{0}$, and remains equal to $\omega _{c}$, from $t=0$ to
%\begin{equation}
%	t_{f}=\pi /\left( 2\omega _{c}\right) ,  
%	\label{T}
%\end{equation}
%and at moment $t=t_{f}$ the frequency instantaneously changes from $\omega_{c}$ to the final value, $\omega_{f}$. From Eqs. (\ref{quench_func}) and (\ref{VA_solution}), we have the following solution: 
\begin{equation}
	a(t)=\sqrt{1+(\omega^2_i-\omega _{c}^{2})\sin^{2}(\omega
		_{c}t)/\omega _{c}^{2}},  
	\label{result}
\end{equation}
%with $\omega _{c}=\sqrt{\omega_i\omega_f}$, 
satisfying the above-mentioned conditions, $a(0)=1$, $a(t_{f})=\gamma $, and $\dot{a}(0)=\dot{a}(t_{f})=0$ \cite{chen2010transient,tangyou2020chaos}.
Thus, Eqs. (\ref{T}) and (\ref{result}) provide a simple exact solution for
the shortcut with just one intermediate frequency, the geometric mean of the initial and final frequencies. Importantly, the solution can be generalized for Eq. (\ref{GEE}), without requiring a explicit form. The energy conservation in this perturbative Kepler problem implies $U(a_{\mathrm{i}})=U(a_{
	\mathrm{f}})$, or, in an explicit form,
\begin{equation}
	\omega _{c}=\sqrt{\frac{A}{a_{\mathrm{i}}^{2}a_{\mathrm{f}}^{2}}+\frac{2B}{a_{\mathrm{i}}a_{\mathrm{f}}(a_{\mathrm{i}}+a_{\mathrm{f}%
		}) }}.  \label{tilde}
\end{equation}
In this case, a simple expression for $t_{f}$ is not available, but it can
be written in the form of an integral:
\begin{equation}
	t_{f}=\int_{a_{\mathrm{i}}}^{a_{\mathrm{f}}}\frac{da}{\sqrt{2\left[ U(a_{
				\mathrm{in}})-U(a)\right] }},  \label{bangt}
\end{equation}
where $a_{\mathrm{i}}$ and $a_{\mathrm{f}}$ are given by Eq. (\ref{Reference}). Accordingly, the trajectory of $a(t) $ can be obtained 
%in a numerical form 
{numerically} from Eq. (\ref{GEE}). 
%with the intermediate $\omega_{c}$ and boundary conditions. 
{The corresponding fidelity is plotted in Fig. \ref{fidelity} along with the results of the frictionless quench. This figure shows that} 
%Fig. \ref{fidelity} illustrates 
the two-jump STA {is less accurate} than the STA by using smooth polynomial function, when the final frequency is smaller.
% \new{[FROM THE FIGURE IT LOOKS THE OPPOSITE].} 
%, yielding lager jump. 
%Here for two-jump trajectory the final times (\ref{bangt}) can be calculated from Eq. (\ref{bangt}), which are different with $\omega_{0}=1$ and various final frequencies $\omega_f$, related to boundary conditions through Eq.  (\ref{Reference}).  
{Another important difference between the two approaches is the fact that in the STA protocol discussed in the previous section the final time is a free parameter, whereas in the present case it is fixed by the trap frequencies and the interaction strength, see Eq. (\ref{bangt}).} 
We should emphasize that the larger repulsive interaction somehow speed up the {two-jump STA} 
%\new{[?? THERE IT WAS $t_{f}=1$]} , 
e.g. 
$t_f = 4.9658$ for $g=0.01$ while $t_f= 4.5380$ for $g=100$, when $\omega_i=1$ and $\omega_{f}=0.1$.

%\subsection{Resonances induced by periodic driving}
%
%The extended parametric resonances firstly proposed in \cite{juan1999prl}, the corresponding spatial resonate induced by a periodic function $\lambda(t) = 1+\epsilon\cos(\omega_Rt)$.  
%Corresponding stability phase diagram which dependent on parameters $\epsilon$ and $\omega_R$, has given in\cite{juan1999prl}.
%In this section, we shall concentrate on the general stability diagram of the system in the arbitrary interaction. 
%For a periodic driving, we rewrite the GEE $(\ref{GEE})$ as
%\beq\label{GEE1}
%\ddot{a} + \lambda(t)a = \frac{A}{a^3}+\frac{B}{a^2}
%\eeq
%
%
%\begin{figure}[tbp]
%\includegraphics[width=\columnwidth]{f99.eps}
%\caption{The resonance induced by a extend periodic driving function $\omega(t) = 1+\epsilon \sin(\omega_f t)$. The corresponding resonance emerges with different interacting strength $g = 0.01, 1, 100$.  We have set $\omega_f = 0.2$, $\epsilon= 0.1$.}
%	\label {fig}
%\end{figur}

 \section{Discussion}
 
  \begin{figure}[tbp]
	\includegraphics[width=\columnwidth]{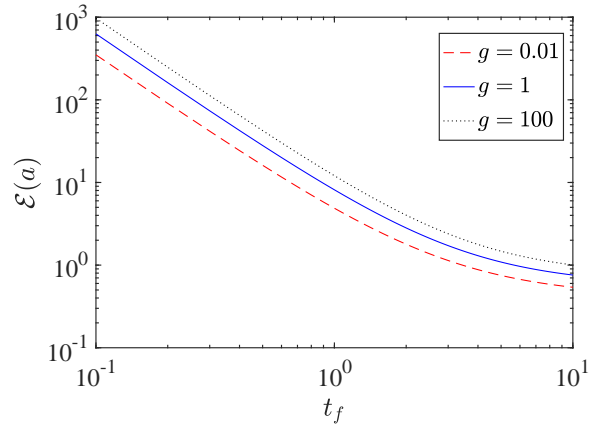}
	\caption{
	Dependence of the 
	{the} time-averaged energy {$\mathcal{E}$ in Eq. (\ref{energycost}),} 
	on the short time $t_f$, where the parameters are the same as those in Fig. \ref{sta}. }
	\label{fig4}
\end{figure}

In the STA {protocol}, there is a trade-off between speed and cost in shortcuts to adiabaticity \cite{AbahNJP19}. In principle, the transient energy excitation of the STA protocol, described by Eq. (\ref{totalE}), is stipulated by the time-energy uncertainty,
which implies an increase of the energy for shorter
times.  In detail,  Fig. \ref{fig4}, illustrates the exponential scaling  $\mathcal{E} \propto 1/t^2_f$ for different regimes of atomic interaction by using the STA protocols in Fig. \ref{sta}, where the time-averaged energy  $\mathcal{E}$ is calculated by Eq. (\ref{energycost}). Moreover, for the same $t_f$, the energy excitation is higher when the interaction is lager. In this sense,   the given energetic cost gives the tight bound on the running time of STA.  Additionally, we can obtain other scaling exponential, i.e. $t_f \propto (\omega_i/\omega_f)^{1/2}$ in the non-interacting limit while  $t_f \propto  (\omega_i/\omega_f)^{3/2}$ in the TF limit \cite{tangyou2020chaos}. 
Instead of using the polynomial and two-jump trajectories, one can further apply the Pontryagin's maximum principle in optimal control theory \cite{Stefanatos2010pra,Stefanatos2012pra} to design the time-optimal STA with the bounded trap frequency. We expect that the atomic interaction slows down the frictionless quenches,  and the minimal time of STA in the intermediate interaction regime is  between the ones in non-interacting and TF limits, see Ref. \cite{tangyou2020chaos}. Moreover, this has the fundamental implications on the quantum speed limit and the third law of thermodynamics in request of absolute zero temperature \cite{rezek2009quantum}.

\section{Conclusion}
 
To conclude, we have employed the effective scaling approach to derive the Ermakov-like equation (\ref{GEE}) for the scaling parameter interpolating between the noninteracting and the TF limits.  By combining inverse engineering with the appropriate boundary conditions, this provides a general way to design accurate STA for 1D Bose gas, for arbitrary values of the atomic interaction strength.  
%Here  we concentrate on  the  1D Bose gas trapped in an ideal cigar-shape optical trap by neglecting the transverse coupling. 
{Here we have considered the case of a quasi-1D condensate confined in cigar-shaped trap, with a tight transverse frequency (that is, much larger than the longitudinal one). These results} 
%However, our results presented here 
can be easily generalized to 
%3D case, see Ref.  \cite{Michele2020PRR}, 
{the 3D case} in different interaction regimes, see Ref.  \cite{Michele2020PRR}. 

In addition,  the effective scaling approach is harnessed to design {a} STA protocol 
%working for the 
{for a} 
trapped 1D Bose gas in arbitrary repulsive interacting regime, where the previous methods, such as dynamical invariant or scaling approach, can not work successfully. We emphasize that the method is similar 
%with
{to}, yet different from, the variational approximation \cite{zoller1996prl,tangyou2020chaos}. In the effective scaling approach, the scaling solution in the hydrodynamic regime is used as ansatz, but 
the Gaussian-shaped ground state in non-interacting limit is replaced as a preassumed ansatz in variational approximation. In this sense,  the effective scaling approach has more reasonable accuracy, when the atomic interaction with the arbitrary value is considered. Finally, some extensions are interesting for further exploration, for instance, {the} soliton dynamics by quenching the interactions of {the} BEC from repulsive to attractive \cite{Excitation2019prl,tangyou2020pra}, or  the expansion of {a} Bose gas in the crossover from TF to TG regimes  \cite{santos2002prl}.

\section*{Acknowledgments}
This work 
%is 
{has been} partially supported 
%from 
{by} NSFC
({Grant No.} 12075145), STCSM ({Grant No.} 2019SHZDZX01-ZX04), and the Program for Eastern Scholar. 
{M.M. and X. C. acknowledge support by the Spanish Ministry of Science and the European Regional Development Fund through PGC2018-101355-B-I00 (MCIU/AEI/FEDER, UE) and the Basque Government through Grant No. IT986-16.
X. C. acknowledges support by the Ramon y Cajal program (Grant No. RYC-2017-22482), the EU FET Open Grant Quromorphic (Grant No. 828826), and  EPIQUS (Grant No. 899368).}
%Basque Government IT986-16, Spanish Government PGC2018-095113-B-I00 (MCIU/AEI/FEDER, UE), EU FET Open Grant Quromorphic (828826) as well as EPIQUS (899368).
T.Y. H.  acknowledges support by the CSC fellowship (202006890071). 

\bibliographystyle{apsrev4-1}
\bibliography{ref}
\end{document}